\documentclass[aps,prl,twocolumn,showpacs,amsmath,amssymb,groupedaddress]{revtex4-1}
\usepackage{graphicx}
\usepackage{dcolumn}
\usepackage{bm}
\usepackage{color}
\usepackage{longtable}
\usepackage{multirow}

\begin{document}

\date{\today}

\title{Charge density wave with anomalous temperature dependence in UPt$_2$Si$_2$}

\author{Jooseop Lee,$^{1,2}$}
\author{Karel Proke\v{s},$^{3}$} 
\author{Sohee Park,$^{1}$} 
\author{Igor Zaliznyak,$^{4}$} 
\author{Sachith Dissanayake,$^{5}$} 
\author{Masaaki Matsuda,$^{5}$} 
\author{Matthias Frontzek,$^{5}$} 
\author{Stanislav Stoupin,$^{2}$} 
\author{Greta L. Chappell$^{6,7}$} 
\author{Ryan E. Baumbach$^{6,7}$}
\author{Changwon Park,$^{1}$} 
\author{John A. Mydosh,$^{8}$} 
\author{Garrett E. Granroth,$^{5}$}
\author{Jacob P. C. Ruff$^{2}$}

\affiliation{$^{1}$CALDES, Institute for Basic Science, Pohang 37673, Republic of Korea}
\affiliation{$^{2}$CHESS, Cornell University, Ithaca, New York 14853, USA}
\affiliation{$^{3}$Helmholtz-Zentrum Berlin f\"{u}r Materialien und Energie, Berlin 14109, Germany}
\affiliation{$^{4}$CMPMSD, Brookhaven National Laboratory, Upton, New York 11973, USA}
\affiliation{$^{5}$Neutron Scattering Division, Oak Ridge National Laboratory, Oak Ridge, Tennessee 37831, USA}
\affiliation{$^{6}$National High Magnetic Field Laboratory, Florida State University, Tallahassee, Florida 32310, USA} 
\affiliation{$^{7}$Department of Physics, Florida State University, Tallahassee, Florida 32306, USA}
\affiliation{$^{8}$Institute Lorentz and Kamerlingh Onnes Laboratory, Leiden University, Leiden 2300 RA , The Netherlands}

\begin{abstract}

Using single crystal neutron and x-ray diffraction, we discovered a charge density wave (CDW) below 320 K, which accounts for the long-sought origin of the heat capacity and resistivity anomalies in UPt$_2$Si$_2$. The modulation wavevector, $\bm{Q}_{\rm{mod}}$, is intriguingly similar to the incommensurate wavevector of URu$_2$Si$_2$. $\bm{Q}_{\rm{mod}}$ shows an unusual temperature dependence, shifting from commensurate to incommensurate position upon cooling and becoming locked at $\approx$ (0.42 0 0) near 180 K. Bulk measurements indicate a cross-over toward a correlated coherent state around the same temperature, suggesting an interplay between the CDW and Kondo-lattice-like coherence before coexisting antiferromagnetic order sets in at $T_{\rm{N}}$ = 35 K.

\end{abstract}

\maketitle

In strongly correlated 5$f$-electron systems, interactions with comparable strength compete due to the extended nature of the wave-functions. This competition leads to a rich variety of exotic states, which can hardly be understood with conventional models from $d$- or 4$f$- electrons physics~\cite{StewartRMP1}. In metallic U-based heavy fermion compounds with strong hybridization effects of the surrounding ligands, an exceptional coexistence of unusual phases occurs as, for example, in the hidden order superconductor URu$_2$Si$_2$. The nature of the $``$hidden order$"$ parameter responsible for the heat capacity anomaly is still debated, more than 30 years after its discovery~\cite{MydoshRMP}.

UPt$_2$Si$_2$ is a closely related intermetallic compound of the U$T_{2}M_{2}$ ($T$: transition metal, $M$: Si or Ge) family whose Pt-5$d$ electrons hybridize with the U-5$f$ states. UPt$_2$Si$_2$ adopts the CaBe$_{2}$Ge$_{2}$ crystal structure and orders magnetically at $T_{\rm{N}}$ = 35 K with wavevector $\bm{Q}_{\rm{M}}$ = (1 0 0), where ferromagnetic (FM) $ab$-planes are stacked antiferromagnetically (AFM) along $c$-axis, with a large ordered magnetic moment of $\approx$ 2 $\mu_B$ \cite{SteemanJPCM,NieuwenhuysPRB,ProkesArxive}. Consequently, UPt$_2$Si$_2$ has long been considered a rare example of a uranium intermetallic compound with localized 5$f$-electrons where magnetism can be explained within a simple crystal-field level scheme~\cite{NieuwenhuysPRB}. Several recent studies~\cite{Grachtrup2012,ElgazzarPRB,Grachtrup2017,LeePRL}, however, questioned the degree of electron localization in this system. High field measurements suggest that phase transitions under applied magnetic field should be understood in terms of Fermi surface effects~\cite{Grachtrup2012}. This approach has been further supported by a density functional theory (DFT) calculation which favors a scenario where the 5$f$-electrons are mostly itinerant~\cite{ElgazzarPRB}. A recent inelastic neutron scattering study reveals a dual nature, both itinerant and local, in the magnetic excitation spectrum of UPt$_2$Si$_2$ at low temperatures~\cite{LeePRL}.

The crystal structure of UPt$_2$Si$_2$ and its relationship with electronic properties have been another important longstanding controversy~\cite{Otop2004,SullowJPSJ,SullowPRL}. The x-ray diffraction measurements found large anisotropic thermal factors on one of the two Pt and Si sites, which were interpreted as an indication of strong crystallographic disorder~\cite{SullowJPSJ}. This purported disorder was thought to be responsible for the highly anisotropic resistivity with Anderson localization along the $\it{c}$-axis.

The origin and the nature of the disorder in this moderately heavy fermion system, however, has long been open to questions. Most U$T_{2}M_{2}$ compounds crystallize into the ThCr$_2$Si$_2$ type structure (space group $I$4$m m m$) and only a few, including UPt$_2$Si$_2$, have been reported to adopt the primitive CaBe$_2$Ge$_2$ type structure. Therefore, when a specific heat anomaly and a concomitant cusp in the electrical resistivity around room temperature were found~\cite{BleckmannJMMM}, it was anticipated that there could be a structural transition between the two closely related crystal structures as in UCo$_2$Ge$_2$ \cite{Endstra91}, leading to disorder. However, x-ray diffraction measurements on a polycrystalline sample did not find any indication of a structural change. Moreover, a comparative electrical transport study of as-grown and annealed samples showed that the effect of the presumed disorder does not depend on synthesis conditions or impurity level~\cite{BleckmannJMMM}.

\begin{figure*}[th]
\includegraphics[width=0.7\hsize]{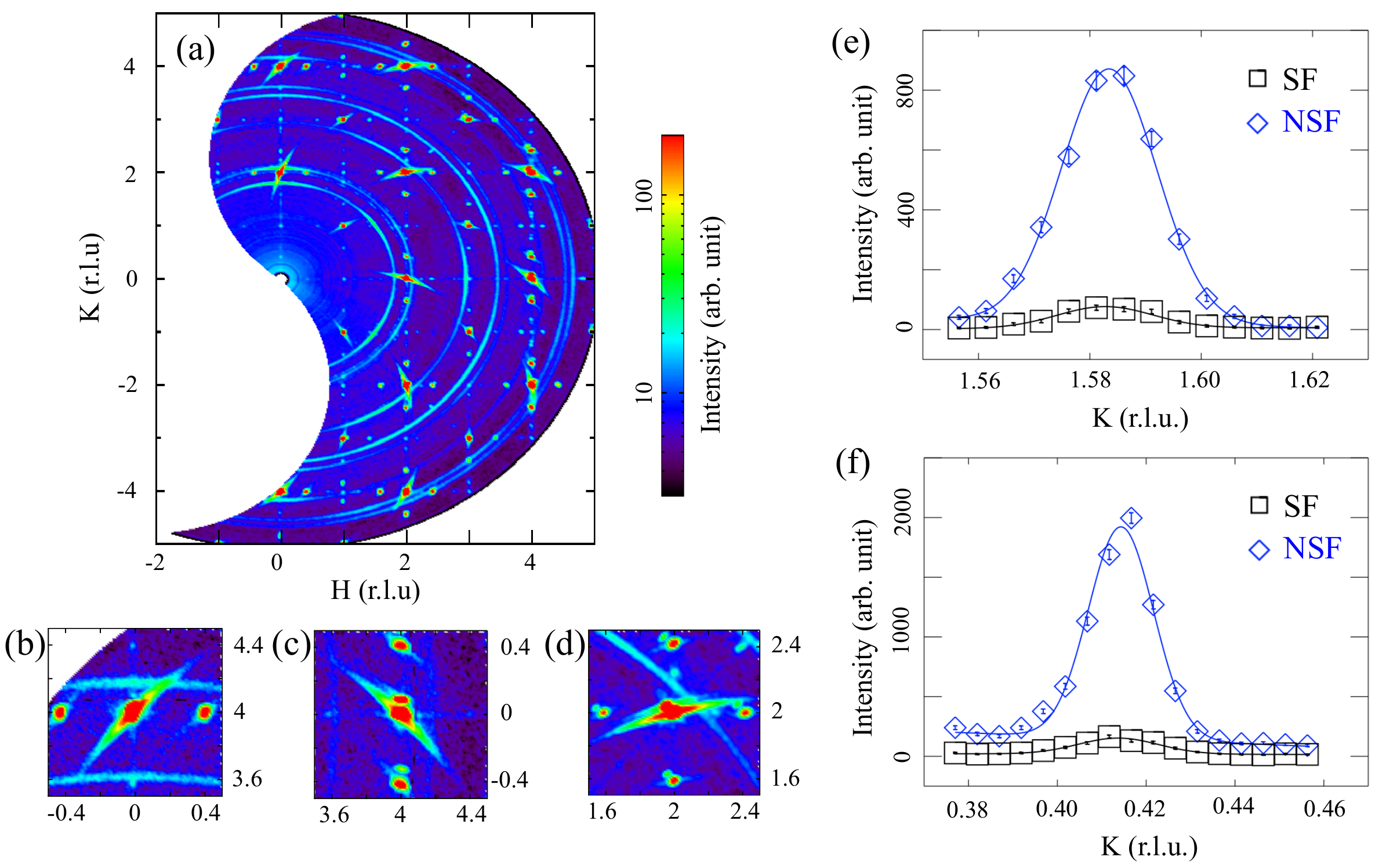}
\centering
\caption{Neutron diffraction measured at 50 K with WAND. (a) a full map in (H K 0) plane and zoomed-in views around (b) (0 4 0), (c) (4 0 0), and (d) (2 2 0) showing peak extinction rules that reveal the polarization of atomic displacements. SF and NSF intensities from PTAX measured at 50 K at (e) $\textbf{Q}$ = (2 0.42 0) and (f) (2 1.58 0), where solid lines are fit to a Gaussian.}
\label{fig:1}
\end{figure*}

Here, we report single crystal neutron and x-ray diffraction measurements, which reveal the discovery of an periodic lattice modulation that most likely results from a charge density wave (CDW) which develops in UPt$_2$Si$_2$ below $T_{s}$ $\approx$ 320 K. The appearance of the CDW is responsible for the heat capacity and electrical resistivity anomalies and resolves the controversy surrounding the putative ``disorder". The CDW wavevector ($\tau$ 0 0) shows an interesting temperature dependence, shifting up on cooling from a commensurate value of $\tau$ = 0.40 just below $T_{s}$ to lock-in value at $\tau$ $\approx$ 0.42 around 180 K. Around the same temperature, electrical transport data indicates a cross-over from a local to a coherent electronic state on cooling, suggesting an interplay between the CDW and Kondo-lattice physics in this system. This unusual type of CDW that involves strongly correlated U-5$f$ states appears at relatively high temperatures and coexists at low temperatures with an antiferromagnetic order. Intriguingly, the incommensurate CDW modulation wavevector, $\bm{Q}_{\rm{mod}}$ = ($\tau$ 0 0) with $\tau$ $\approx$ 0.4, resembles the incommensurate wavevector of URu$_2$Si$_2$ where the softening of magnetic excitation is observed ~\cite{BareilleNatComm,ElgazzarNatMat}, implying that similar underlying physics may be at play. 

Neutron diffraction measurements were carried out at the Wide Angle Neutron Diffractometer (WAND) and HB1 Polarized Triple-Axis Spectrometer (PTAX) at the High Flux Isotope Reactor (HFIR), ORNL and E4 two-axis diffractometer~\cite{E4} at the Helmholtz-Zentrum Berlin. For the PTAX measurement, the neutron spin was aligned parallel to wavevector transfer, $\bm{q}$, so that nuclear and magnetic scattering were measured in Non-Spin-Flip (NSF) and Spin-Flip (SF) channel, respectively. The flipping ratio of PTAX was approximately 15. Synchrotron x-ray experiments were performed at the A1 beamline at Cornell High Energy Synchrotron Source (CHESS). We carried out density-functional theory (DFT) calculations via the Generalized Gradient Approximation (GGA) using Perdew-Burke-Ernzerhof (PBE) functional~\cite{DFT1}, as implemented in the VASP package~\cite{DFT2,SuppleExp}. 

\begin{figure*}[th]
\includegraphics[width=0.85\hsize]{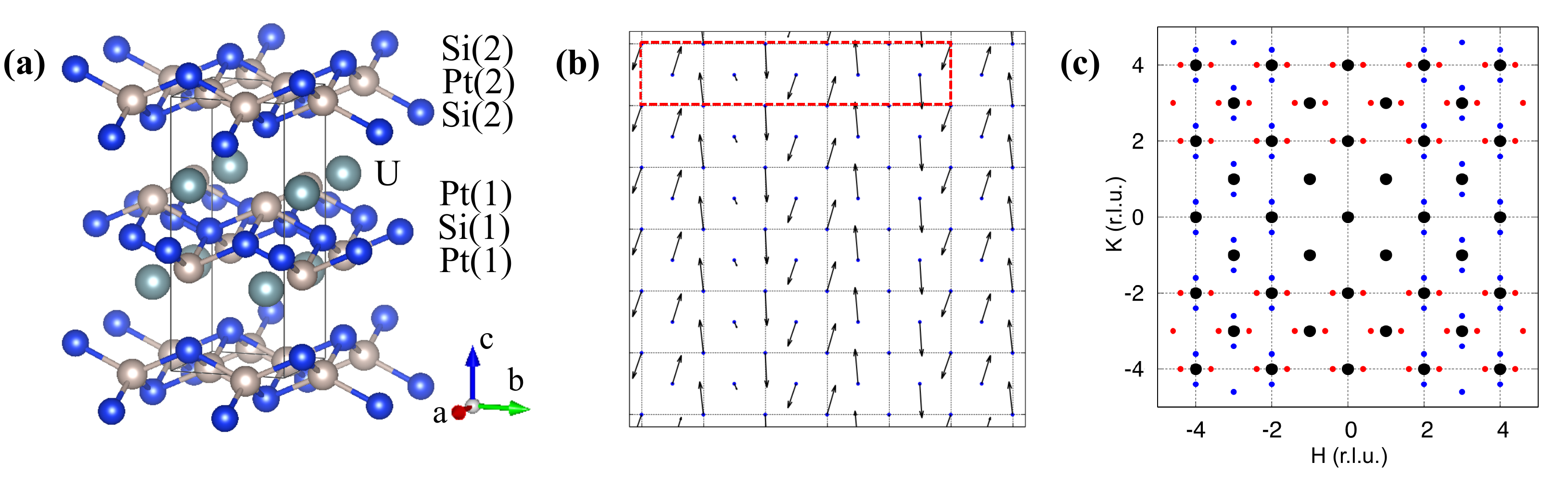}
\centering
\caption{(a) Crystal structure of UPt$_2$Si$_2$ where bonds between Pt and Si atoms illustrate two inequivalent Pt-Si layers in CaBe$_2$Ge$_2$ structure. The atomic labels follow the convention of \cite{SullowJPSJ}, origin 2 of $P$4/nmm, which reported  disorder in UPt$_2$Si$_2$. (b) Calculated displacements of Pt(2) atoms in the ab-plane within one domain. The red rectangle is a 5$\times$1$\times$1 calculation supercell, where blue dots and black arrow represent the original position and 15 times of atomic displacement, respectively. (c) Simulated reflection pattern from the displaced Pt atoms. The radii of the dots are proportional to the squares of Fourier components while Bragg peaks are reduced by 30\% for clarity.}
\label{fig:2}
\end{figure*}

Figure~\ref{fig:1}~(a) presents single crystal neutron diffraction data obtained at the WAND neutron diffractometer at T = 50 K, which is above T$_{\rm{N}}$ = 35 K but well below the reported heat capacity anomaly temperature of $\approx$~320 K~\cite{BleckmannJMMM}. We observe a pattern of satellite peaks offset from the original P4/nmm Bragg reflections by a wavevector $\bm{Q}_{\rm{mod}}$ = ($\tau$ 0 0) or (0 $\tau$ 0), where $\tau$ $\approx$~0.42. We investigated the origin of these superlattice reflections by performing neutron polarization analysis on the (2 0.42 0) and (2 1.58 0) satellite peaks at 50 K at HB1 triple-axis spectrometer as shown in Fig.~\ref{fig:1}~(e) and (f), respectively. NSF scattering dominates the intensity of both reflections, confirming their non-magnetic, charge/lattice origin. Our findings thus correct the previous reports on the atomic disorder~\cite{SullowJPSJ}, which in fact resulted from the mis-assignment of the observed changes of the lattice principal Bragg reflection intensities to disorder rather than lattice modulation induced by CDW. 

By virtue of the WAND diffraction patterns covering wide reciprocal space area, Fig.~\ref{fig:1}~(a), we can clearly identify the polarization of the observed atomic displacements. As shown in Figs.~\ref{fig:1} (b) and (c), the intensities of longitudinal superlattice peaks, (H~$\pm\tau$ 0 0), or, equivalently, (0 K~$\pm\tau$ 0), are absent, or very weak compared to those of transverse superlattice peaks, ($\pm\tau$ K 0) and (H $\pm\tau$ 0). For example, the integrated intensity of a longitudinal (4.42 0 0) peak is about 240 times smaller compared to the corresponding transverse (4 0.42 0) peak. For lattice Bragg peaks near the diagonal (H H 0) direction we find all 4 satellite reflections with comparable intensity, Fig.~\ref{fig:1}~(d). There exist weak reflections at (H K 0) with H + K = odd inconsistent with P4/nmm symmetry, which, however, turns out to be due to an internal strain~\cite{ProkesArxive}.

The intensity of a satellite peak at a momentum transfer $\bm{q}$ can be expressed as~\cite{KBLEE,IgorPRB},
\begin{equation}
    I(\bm{q}) \approx \Big|\sum_{\nu} (\bm{q} \cdot \bm{\epsilon}_{\nu}) f_{\nu}(\bm{q} \pm \bm{Q}_{\rm{mod}}) \Big|^2 \delta(\bm{q}-\bm{G} \pm \bm{Q}_{\rm{mod}}) ,
\label{eq:eq1}
\end{equation}
e.g. using Taylor expansion of Jacobi-Anger identity. Here, $\bm{\epsilon}_{\nu}$ is the displacement of atom ${\nu}$ in the unit cell, $f_{\nu}$ is the partial structure factor of atom ${\nu}$, and $\bm{G}$ is the reciprocal lattice vector. The $(\bm{q} \cdot \bm{\epsilon}_{\nu})$ prefactor in the satellite reflection intensity is responsible for the observed polarization dependence of the diffraction pattern, indicating that the atomic displacements, $\bm{\epsilon}_{\nu}$, in UPt$_2$Si$_2$ are mostly transverse, either along the $\it{a}$-axis direction for $\bm{Q}_{\rm{mod}}$ $\approx$ (0 0.4 0), or along the $\it{b}$-axis direction for $\bm{Q}_{\rm{mod}}$ $\approx$ (0.4 0 0). While the observed pattern of superlattice reflections can be regarded as overlaid patterns from two $\bm{Q}$-domains, within our resolution we did not detect any clear sign of tetragonal symmetry breaking in the main Bragg reflections.

\begin{figure}[tp]
\includegraphics[width=0.70\hsize]{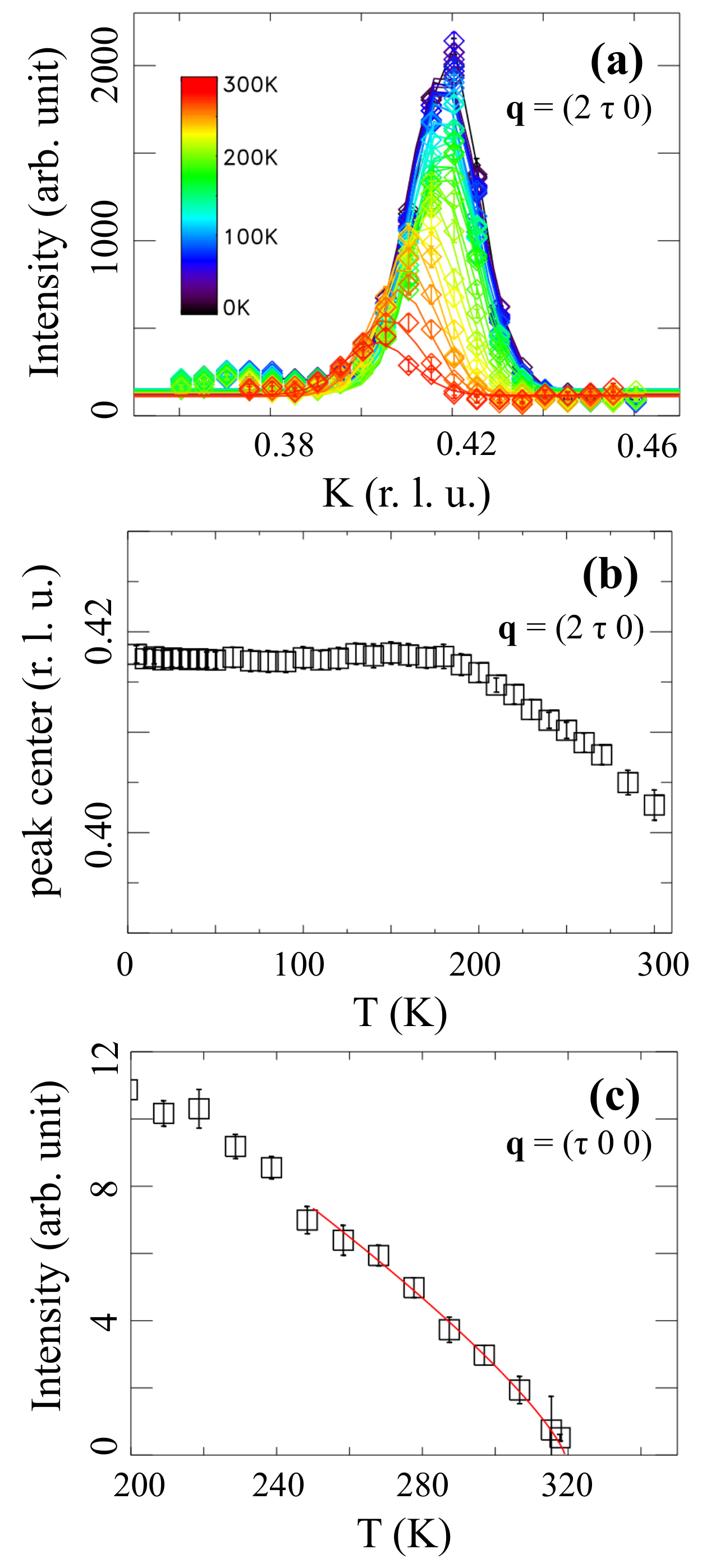}
\centering
\caption{(a) Temperature dependence of the (2 $\tau$ 0) superlattice peak. Solid lines are fit to Gaussian. (b) Temperature dependent peak position of $\tau$, which locks in around T$_{coh}$ = 180 K. (c) Integrated intensity of ($\tau$ 0 0) measured up to 317 K. Red solid line is the best fit to critical behavior described in the text. The data presented in panels (a) and (b) are measured from 4 K to 300 K at PTAX in NSF configuration and therefore are of charge/lattice origin in nature. The unpolarized data in (c) were measured at E4.}
\label{fig:3}
\end{figure}

To understand the atomic modulation pattern in real space, we performed DFT calculation by relaxing the atomic positions in a 4$\times$1$\times$1, 5$\times$1$\times$1, and 6$\times$1$\times$1 supercell. For the 5$\times$1$\times$1, case, we find a reduction of the total energy by 8.6 meV per unit cell with the presence of the CDW. This confirms the CDW to be unidirectional, that is, each Q-domain is described by a single wave vector. As shown in Fig.~\ref{fig:2} (a), there are two kinds of Pt-Si layers in UPt$_2$Si$_2$. In the Si(2)-Pt(2)-Si(2) layer, a square lattice of Pt atoms is sandwiched between Si atoms, while in the Pt(1)-Si(1)-Pt(1) layer the situation is inverted. Our calculation shows that about 90\% of displacement is in the $ab$-plane of the Si(2)-Pt(2)-Si(2) layer, in agreement with previously reported putative disorder~\cite{SullowJPSJ}. It is interesting that the lattice modulation of sandwiched metallic layers is quite ubiquitous~\cite{TMDC1} as in the Pt-based layered superconductor LaPt$_2$Si$_2$~\cite{CDWSooranKim,CDWUeda,CDWFalkowski}. It should be, however, noted that the period of the LaPt$_2$Si$_2$ superstructure is close to 3 unit cells compared to the $\approx$ 5 unit cell periodicity in UPt$_2$Si$_2$. Here, we expect the strongly correlated 5$f$ electron states to participate in the bonding due to hybridization between U-5$f$ and Pt-5$d$ states.

The calculated real space displacement pattern of Pt(2) layer is shown in Fig.~\ref{fig:2} (b), where the dashed red rectangle represents the calculated supercell. When the CDW modulation wavevector is given along the $\it{a}$-axis, the atomic displacement is almost along the $\it{b}$-axis. Figure~\ref{fig:2} (c) shows the Fourier transform of the Pt atomic positions, which can well explain the polarization of the observed reflections. Black, red, and blue dots correspond to principal Bragg peaks, satellites peaks from Fig.~\ref{fig:2} (b) structure and its 90$^{\circ}$-rotated domain, respectively.

Figure~\ref{fig:3} (a) presents the evolution of the superlattice Bragg peak at $\bm{q}$ = (2 $\tau$ 0) with temperature. The intensity, Fig.~\ref{fig:3} (c), keeps increasing as the sample is cooled down, indicating gradual development of the superlattice modulation. Here, we find a second-order-like transition with a transition temperature T$_{\rm{s}}$ = 319(8) K, consistent with the previously reported heat capacity and electrical resistivity anomaly temperature~\cite{BleckmannJMMM}. Fitting the temperature dependence of the order parameter to expression $I \propto (T_{c}-T)^{2\beta}$, we obtain the critical exponent $\beta$ = 0.39(10) consistent with the mean-field behavior. Below $T_{\rm{N}}$ = 35 K, the CDW coexists with the antiferromagnetic order. 

Surprisingly, as shown in Fig.~\ref{fig:3}~(b), the satellite peak moves away from the nearly commensurate position of $\tau$ = 0.400(5) observed around 300 K and becomes locked-in at an incommensurate value 0.418(2) below $\approx$ 180 K suggesting an unusual thermal evolution of the CDW. As the energy to lock-in to the commensurate lattice is a function of the CDW amplitude, which increases at low temperature, an opposite, commensurate to incommensurate lock-in of CDW upon cooling observed here is extremely rare~\cite{TokuraPRL,DeanPRX}. Such a CDW wavevector shift suggests that the electronic structure of Uranium plays an important role at low temperatures. 

Previously reported bulk measurements indicate that around the same temperature the $\it{a}$-axis resistivity shows a broad maximum~\cite{AmitsukaPhysB, SuppleBulk} and the magnetic susceptibility begins to deviate from the Curie-Weiss law~\cite{NieuwenhuysPRB}. These observations suggest an onset of a ``Kondo-lattice type" coherence that marks a crossover from local moments and weakly correlated electrons to correlated, heavy-electron behavior. Therefore, the unusual gradual change of $\bm{Q}_{\rm{mod}}$ from commensurate to incommensurate is likely related to a transition from local-moments towards a system with a dual character of the 5$f$ electrons~\cite{LeePRL} facilitated by the 5$f$-5$d$ hybridization and indicates an interplay between CDW and Kondo-lattice like coherence in UPt$_2$Si$_2$. While there are a few systems that show the mere coexistence of CDW and Kondo states~\cite{StewartRMP1,KondoPRB05,KondoSciRep17}, we suggest UPt$_2$Si$_2$ as the first example of coupling between CDW and Kondo-lattice coherence effects.

\begin{figure}[tp]
\includegraphics[width=0.85\hsize]{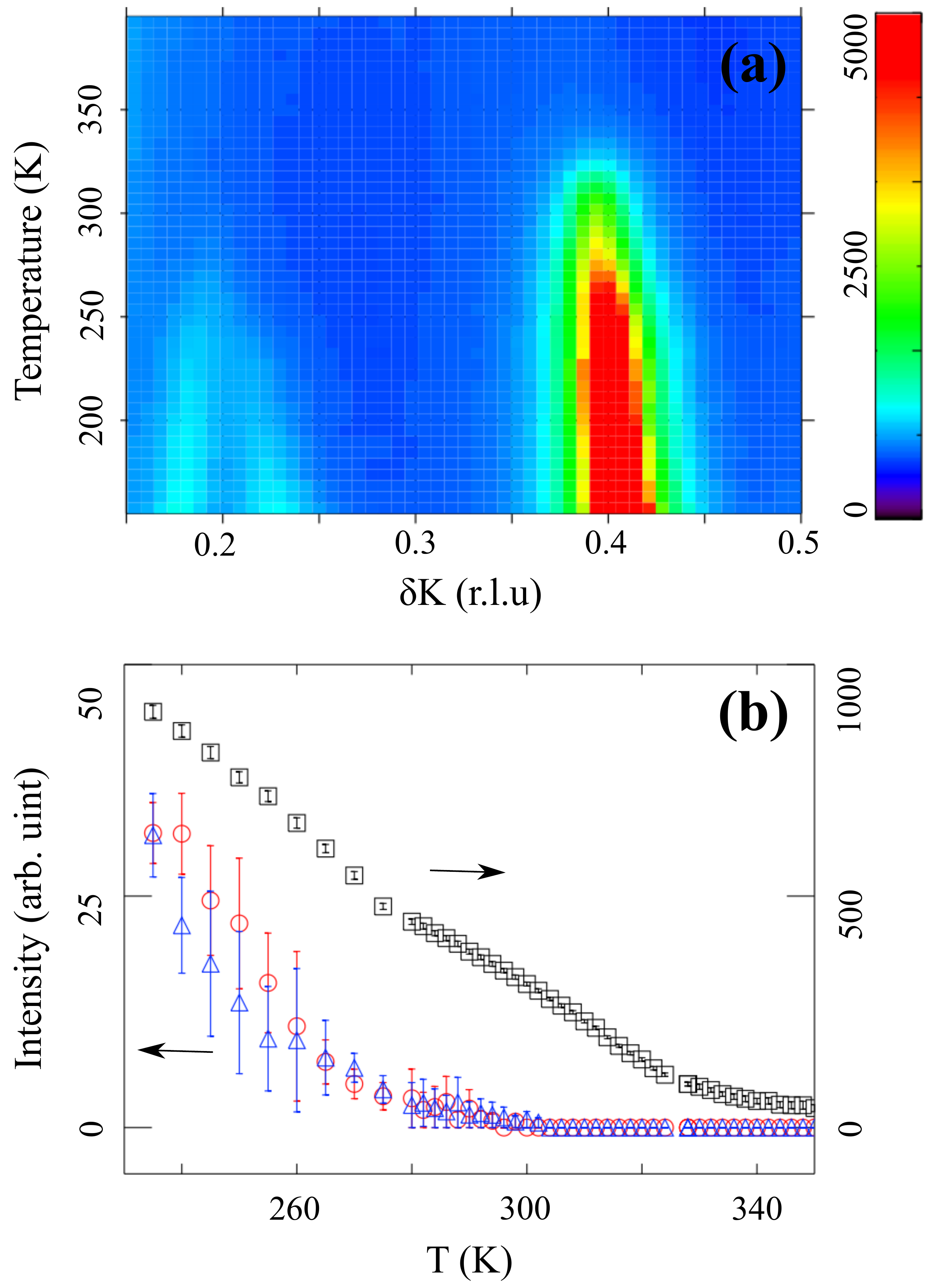}
\centering
\caption{Temperature dependence of the non-resonant synchrotron x-ray diffraction pattern measured using synchrotron x-ray at A1, CHESS. (a) temperature dependence of superlattice harmonics around $q$ = (4 3-$\delta$K 2). (b) temperature dependence of integrated intensities of principal harmonic (black squares), second harmonic (red circles), and third harmonic (blue triangles). The axis for the principal harmonic is shown on the right.}
\label{fig:4}
\end{figure}

The nature of the CDW has been investigated in more detail with a synchrotron x-ray diffraction measurement. A careful inspection of the diffraction pattern reveals that besides the first harmonic superlattice reflections with $\bm{Q}_{\rm{mod}}$, there are also weak superlattice reflections indexable as the second and the third harmonics located around $q$ = (4 3-$\delta$K 2), Fig.~\ref{fig:4} (a). At high temperature, weak second and third harmonics nearly overlap at $q$ = (4 3.8 2) as higher harmonics are found at (4 3+2$\bm{Q}_{\rm{mod}}$ 2) and (4 5-3$\bm{Q}_{\rm{mod}}$ 2), while $\bm{Q}_{\rm{mod}}$ is commensurate, with $\tau$ = 0.4. As the temperature decreases, the second and third harmonic peaks get stronger in intensity and split apart as the incommensurability increases. Below T$_{coh}$, their positions lock-in at the corresponding incommensurate values, 2$\tau$ and 3$\tau$, confirming their higher harmonics origin.

In Fig.~\ref{fig:4} (b), we show the temperature dependence of the integrated intensities of the first three harmonics. The onset of the principal harmonics at T$_{\rm{s}}$ precedes a long tail of critical fluctuation precursor to the CDW at T $>$ T$_{\rm{s}}$, which is emphasized in an energy-integrating measurement without an analyzer. There is also a hint of a second phase transition around 270 K, where the intensities of higher harmonics begin to increase, suggesting that the CDW order parameter becomes two-component, with two displacement polarizations. The existence of higher harmonics indicates that the atomic CDW modulation is not purely sinusoidal, as is also noted in DFT calculation.

It is provocative to compare UPt$_{2}$Si$_{2}$ with URu$_{2}$Si$_{2}$. While these systems seem to be very different at the first glance, one with mainly itinerant and the other with mainly local 5$f$ electron states, they indeed share many common features. They have closely related crystal structures and the same magnetic structures although the magnetic order in URu$_{2}$Si$_{2}$ seems to hover in the background, ready to appear under pressure, dilution, or magnetic field~\cite{AmitsukaPRL,KnafoNatComm}. Thermoelectric measurements~\cite{JohannsenPRB} suggest that both systems show Fermi surface reconstruction with a partial gap opening in hidden order or antiferromagnetic states, respectively~\cite{SuppleCompare}.

Here, we find that both systems have two similar important wavevectors at play~\cite{SuppleWavevectors}. In URu$_{2}$Si$_{2}$, strong and coherent magnetic excitations are observed around the antiferromagnetic wavevector, $\bm{Q}_{\rm{0}}$ = (1 0 0), and the incommensurate wavevector, $\bm{Q}_{\rm{1}}$ $\approx$ (1$\pm$0.4 0 0)~\cite{BroholmPRL,WiebeNatPhy}, whose origin is commonly believed to be Fermi surface nesting. UPt$_{2}$Si$_{2}$, as we find here, develops a CDW with $\bm{Q}_{\rm{mod}}$ = (0.4 0 0) and antiferromagnetic order with $\bm{Q}_{\rm{M}}$ = (1 0 0). Very recent STM measurement shows Moir\'{e} pattern in URu$_2$Si$_2$ possibly from a CDW related to $\bm{Q}_{\rm{1}}$~\cite{arXiv}. It is tempting to suggest that similar wavevectors characterizing the respective ground states in URu$_{2}$Si$_{2}$ and UPt$_{2}$Si$_{2}$ might be more than just a pure coincidence, demanding further attention.


In conclusion, our diffraction experiments demonstrate that a CDW appears near room temperature in a strongly correlated U-based intermetallic compound, UPt$_{2}$Si$_{2}$. In the ground state, this CDW coexists with an AFM order, which appears at a much lower temperature. Higher-harmonic CDW reflections, which arise on cooling below $\approx$ 270 K indicate non-sinusoidal lattice modulation that is also supported by DFT calculations. Furthermore, the CDW wavevector exhibits unique temperature dependence. The commensurate-incommensurate transition on cooling and lock-in near the coherence temperature challenges our current understanding of CDW phenomena and requires taking into account the 5$f$-ligand hybridization. This suggests Kondo-lattice type physics and the dual nature of U 5$f$-electrons at play in UPt$_{2}$Si$_{2}$. The similarity of wavevectors characterizing electronic order in the two seemingly different systems, URu$_{2}$Si$_{2}$ and UPt$_{2}$Si$_{2}$, requires further studies, which might shed light on our understanding of the hidden order phase and unconventional superconductivity.


\begin{acknowledgments}
We thank K. -B. Lee and D. Y. Kim for helpful discussion. The sample synthesis was supported by Stefan S\"{u}llow whose work has been supported by the DFG under contract No. SU229/1. Work at Brookhaven National Laboratory was supported by Office of Basic Energy Sciences (BES), Division of Materials Sciences and Engineering, U.S. Department of Energy (DOE), under contract DE-SC0012704. A portion of this research used resources at the High Flux Isotope Reactor, a DOE Office of Science User Facility operated by the Oak Ridge National Laboratory. Cornell High Energy Synchrotron Source was supported by the NSF and NIH/National Institute of General Medical Sciences via NSF award no. DMR-1332208.
\end{acknowledgments}


\end{document}